\documentstyle[prl,aps,multicol,epsf]{revtex}

\begin{document}

\title{Revealing Superfluid--Mott-Insulator Transition 
in an  Optical Lattice}

\author{V.A.~Kashurnikov$^{1}$, N.V.~Prokof'ev$^{2}$, and 
B.V.~Svistunov $^{2,3}$}

\address{
$^1$Moscow State Engineering Physics Institute, 115409 Moscow, Russia \\
$^2$ Department of Physics, University of Massachusetts, 
Amherst, MA 01003, USA \\
$^3$ Russian Research Center ``Kurchatov Institute", 123182 Moscow, 
Russia }

\maketitle
\begin{abstract}
We study (by an exact numerical scheme) the single-particle density matrix
of $\sim 10^3$ ultracold atoms in an optical lattice with a parabolic
confining potential. Our simulation is directly relevant to the 
interpretation and further development of the recent pioneering 
experiment \cite{Greiner}. In particular, we show that restructuring 
of the spatial distribution of the superfluid component when
a domain of Mott-insulator phase appears in the system, results in a 
fine structure of the particle momentum distribution. This feature may be used
to locate the point of the superfluid--Mott-insulator transition.
\end{abstract}


\begin{multicols}{2}
\narrowtext

The fascinating physics of the superfluid--insulator transition
in a system of interacting bosons 
on a lattice has been attracting constant interest of theorists 
during recent years \cite{Fisher,Sheshadri,FM,Batrouni,Niyaz,PS,Hebert,Schmid}.
Lattice bosons is one of the simplest many-body problems with strong competition 
between the potential and kinetic energy, and a typical example of the 
quantum phase transition system.
One of its great advantages is the 
possibility to study it by powerful Monte Carlo methods which nowadays 
allow simulations of many thousands of particles at low temperature 
with unprecedented accuracy (see, e.g., \cite{Schmid}).
However, until very recently the canonical Bose-Hubbard model
\begin{equation}
H = -t\sum_{<ij>} a^{\dagger}_i 
a^{ }_j +  {U \over 2}\sum_i n_i^2 -\sum_i \mu_i \, n_i\; ,
\label{H}
\end{equation} 
(where $a^{\dagger}_i$ creates a particle on the site $i$, 
$< \! ij \! >$ stands for the nearest-neighbor 
sites, $n_i=a^{\dagger}_i a^{ }_i $, and $t$, $U$, and $\mu_i$, are 
the hopping amplitude, the on-site interaction, and the on-site external field,
respectively) was not particularly useful in the analysis
of realistic systems. The situation has changed with the exciting success
of the experiment by Greiner {\it et al}. \onlinecite{Greiner}
(originally proposed by Jaksch {\it et al}. \cite{Jaksch}) in which
a gas of ultracold $^{87}$Rb atoms was trapped in a 
three-dimensional, simple-cubic optical lattice potential. 
The uniqueness of the new system is that it is adequately described by
the Bose-Hubbard Hamiltonian \cite{Jaksch,Greiner}, and allows virtually
an unlimited control over the strength of the effective interparticle 
interaction $U/t$ and particle density.

The characteristic feature of the experimental setup of Ref.~\cite{Greiner}
is the presence of the overall parabolic potential $V(r)$ which confines the 
sample. This feature could be of great advantage if 
one was able to directly measure the spatial density distribution in the trap.
We recall the structure of the $\mu - U/t$ phase diagram for the 
Bose-Hubbard system \cite{Fisher} which predicts commensurate particle density
distribution for the insulating phase whenever the chemical potential lies
within the Mott-Hubbard gap. The slowly varying (at the length scale of the lattice period)
trapping potential effectively
provides a scan over $\mu$ of this phase diagram at a 
fixed value of $U/t$. 
 
Unfortunately, what is measured in the experiment is not the original spatial density
distribution in the trap, but the absorption image of the free evolving atomic cloud,
after the trapping/optical potential is removed; i.e. the quantity which 
is directly related to the the single-particle density matrix in {\it momentum}
space, $\rho_{\bf kk}=n_{\bf k}$. [This statement implies that in the free evolving 
atomic cloud the interparticle interaction can be neglected; see the 
discussion below.] 
Now, in terms of $n_{\bf k}$ the inhomogeneous
trapping potential is a disadvantage since it broadens the superfluid 
$\delta$-functional contribution at ${\bf k} =0$, and the observed picture 
is a convolution of the original real-space density matrix 
$\rho({\bf r},{\bf r}')$.
As we show below one has then to look at the fine-structure of the 
central peak in the experimental data to decipher the Mott-Hubbard phase diagram
(the ``fading'' of the Bragg peaks in the experiment has very little to do with 
the superfluid--Mott-insulator transitions and happens when the system is already
deep in the insulating phase).      
  
In this Letter we relate quantitatively the particle distribution
in momentum space observed in experiments to the corresponding spatial density
distribution in the trap. Our ultimate goal is to reveal which features (if any)
in the structure of $n_{\bf k}$ indicate unambiguously the presence of the
Mott phase. To this end we perform quantum Monte Carlo simulations of the 
single-particle density matrix for the Bose-Hubbard systems with up to $16^3$ 
lattice sites using continuous-time Worm algorithm \cite{Worm}. We find
that the onset of the phase transition in the trap center should result 
in appearance of at least one satellite peak in $n_{\bf k}$, reflecting a shell-type
form of the superfluid component.
This peak was not mentioned in the experiments of Ref.~\onlinecite{Greiner}. 
We suggest a possible explanation to this fact, and argue that by collimating 
the expanding atomic cloud one can render this peak observable. We also
discuss the role of self-repulsion in the free expanding cloud, which
can affect the simple interpretation of the absorption images in terms
of the initial single-particle momentum distribution.
\end{multicols}

\begin{multicols}{0}
\onecolumn
\vspace*{-1.cm}
\begin{figure}
\begin{center}
\epsfxsize=0.33\textwidth
\hspace*{0.0cm} \epsfbox{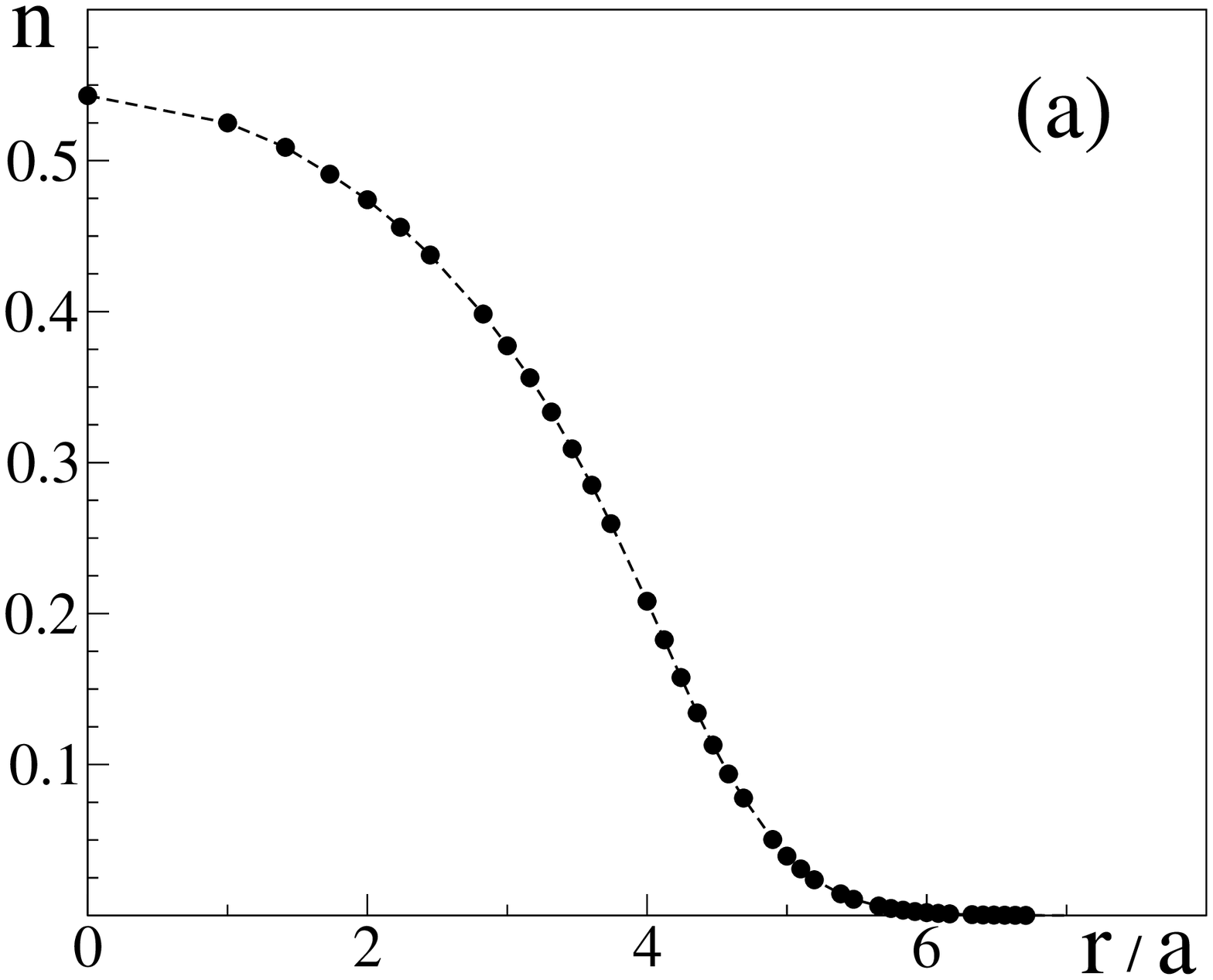}
\epsfxsize=0.33\textwidth
\hspace*{0.0cm} \epsfbox{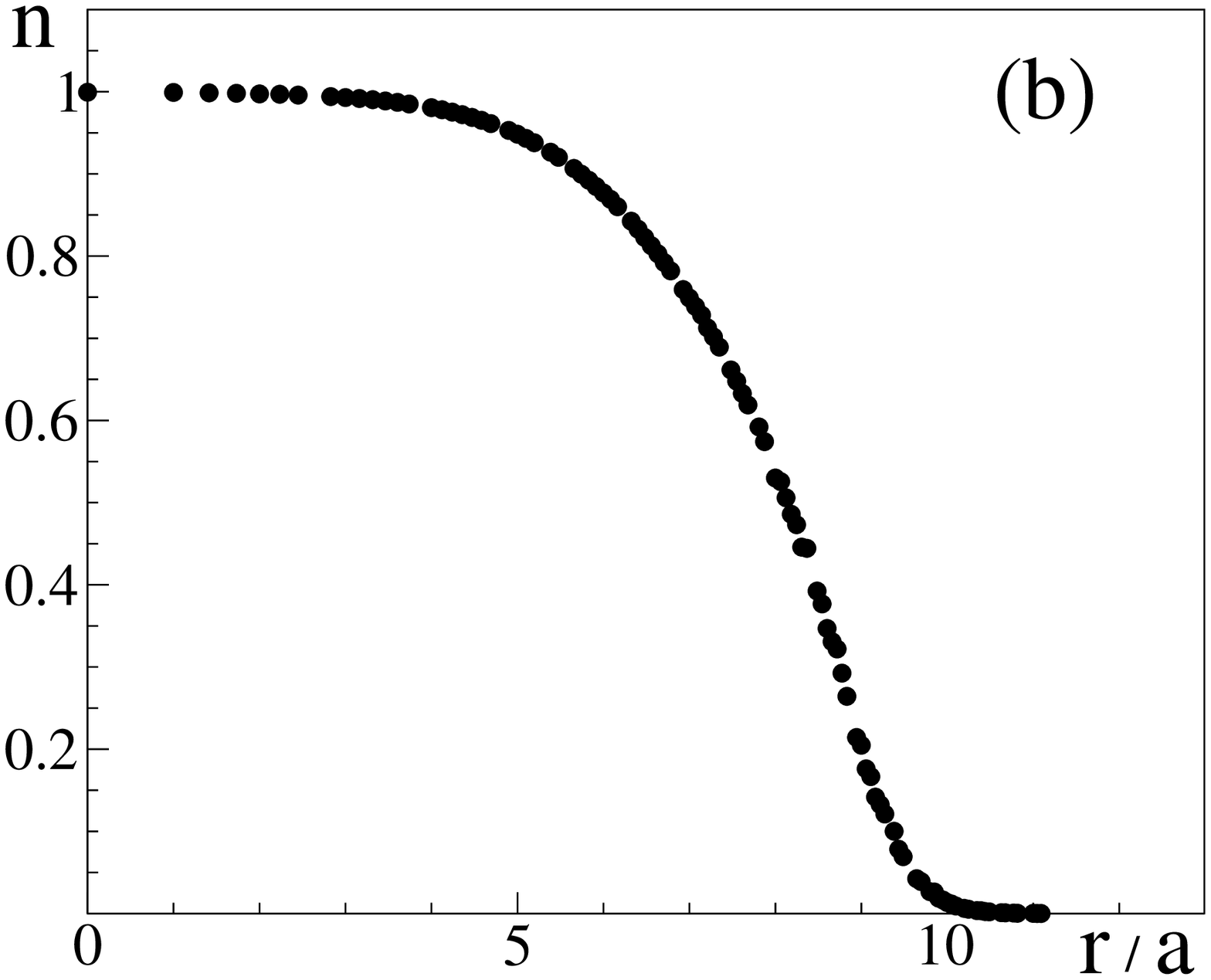}
\epsfxsize=0.33\textwidth
\hspace*{0.0cm} \epsfbox{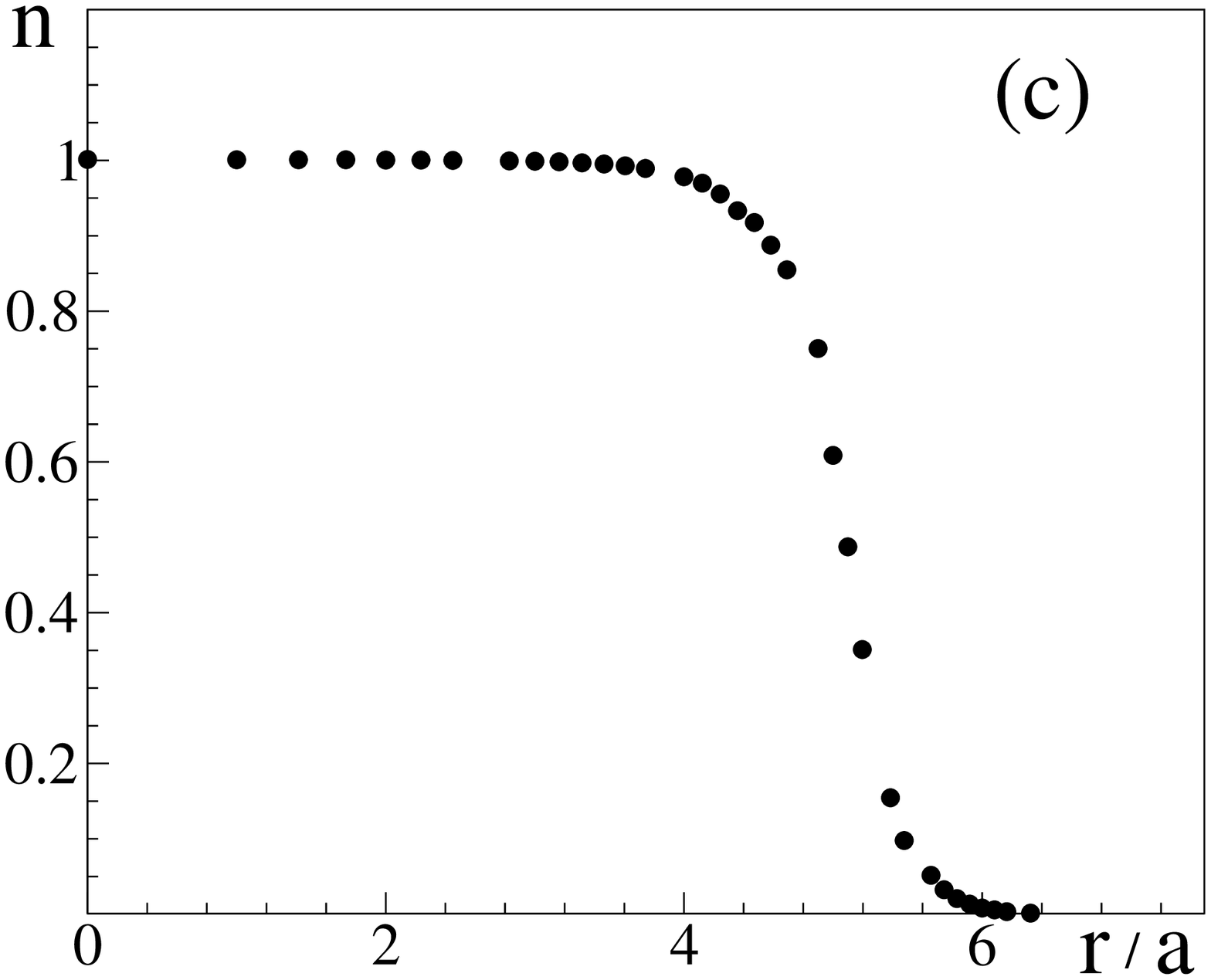} 

\vspace*{-2.5cm}
\epsfxsize=0.33\textwidth
\hspace*{0.0cm} \epsfbox{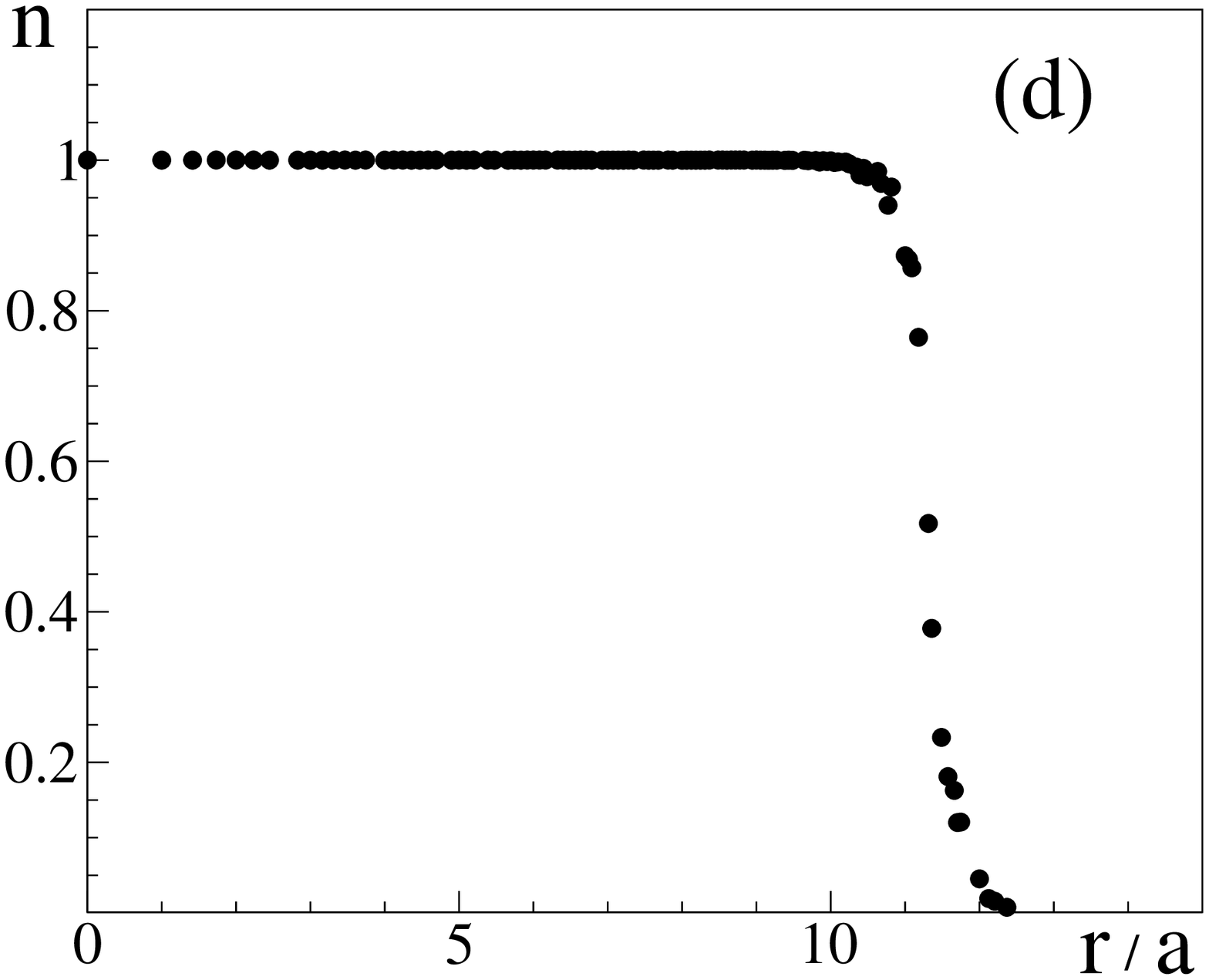}
\epsfxsize=0.33\textwidth
\hspace*{0.0cm} \epsfbox{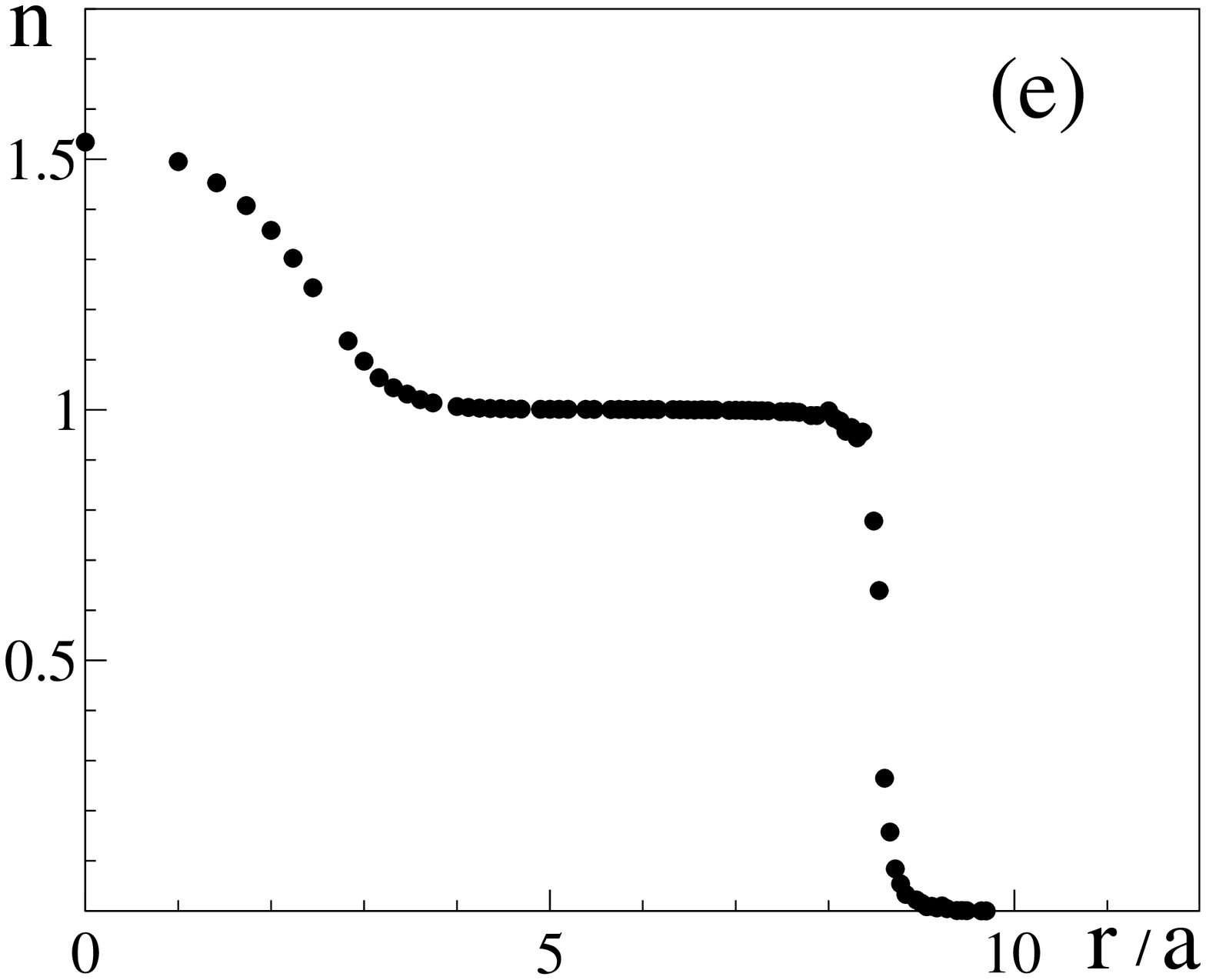}
\epsfxsize=0.33\textwidth
\hspace*{0.0cm} \epsfbox{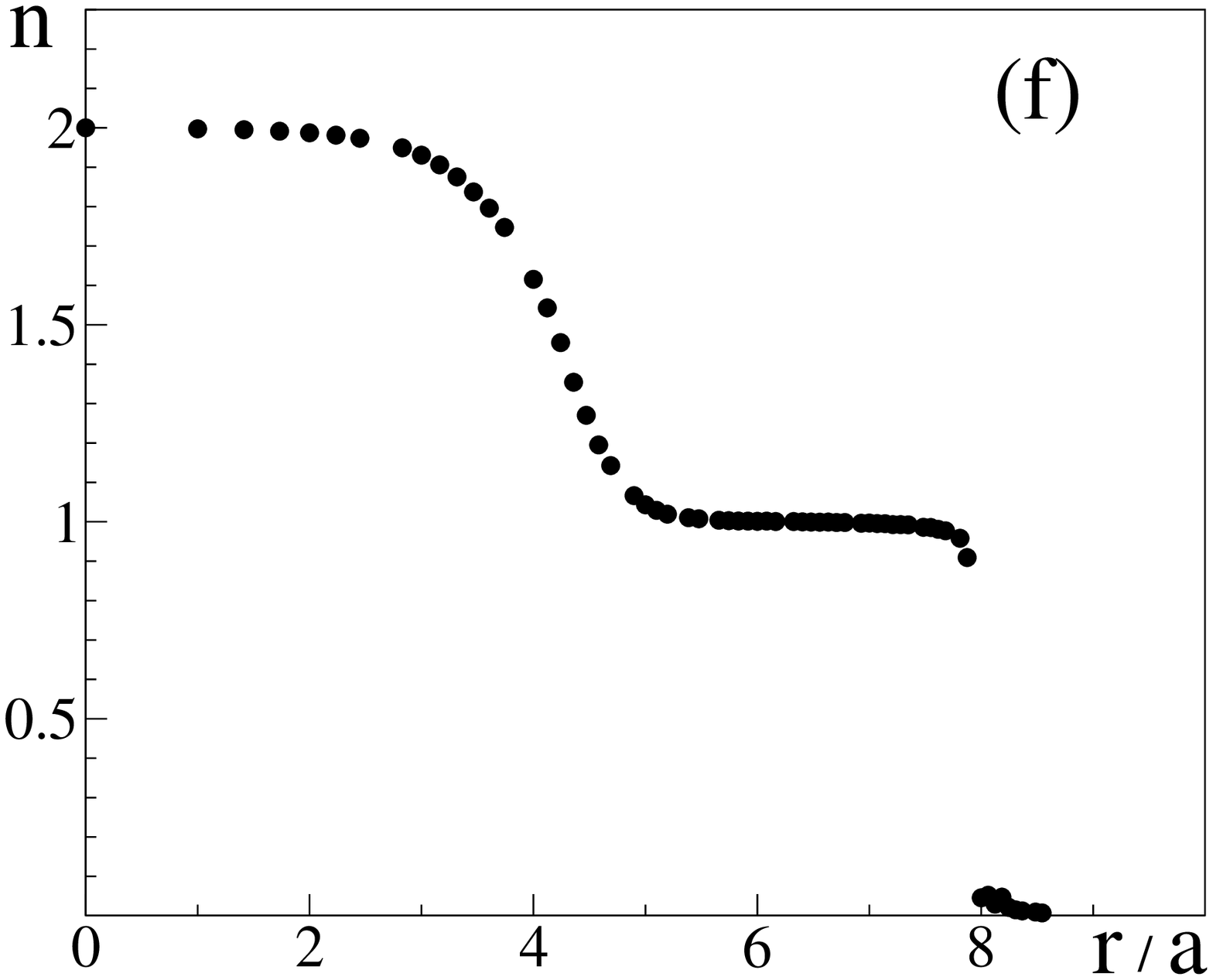}
\end{center}
\vspace*{-1.2cm}
\caption{Particle density distributions (on-site filling factors) 
as functions of the lattice site distance from the trap center
for various coupling parameters and filling factors in the 
center:  $U/t=24$ (a),  $U/t=32$ (b),  $U/t=80$ (c), 
 $U/t=80$ (d),  $U/t=80$ (e),  $U/t=80$ (f).}
 
\end{figure}
\end{multicols}

\begin{multicols}{2} 
In Figs.~1(a-f) we present our data for the density distribution 
as function of the lattice cite distance from the trap center, $r/a$, 
where $a$ is the lattice constant. For all practical purposes one may assume the 
zero-temperature limit here. The simulation was done at finite but very
low $T = 0.2 t$; the relevant energy parameters in this model are the bandwidth
$W=12t$ and $(U/t)_c \sim 35$ \cite{Sheshadri,FM}. In accordance with the phase diagram 
of Ref.~\onlinecite{Fisher}, we observe a shell-type structure of the particle
density with the Mott-insulator phases visible as integer plateau regions.

Next, we relate each of the above figures to the corresponding momentum distribution 
function $n_{\bf k}$. By definition, $n_{\bf k}=\int {\rm d}^3 r {\rm d}^3 r'
\exp [i {\bf k} ({\bf r} - {\bf r}') ] \rho({\bf r},{\bf r}')$, where
$\rho({\bf r},{\bf r}') =
\langle \psi^{\dagger}({\bf r}) \psi({\bf r}') \rangle$, and $\psi({\bf r})$ is 
the bosonic field operator. 
In our case of a single-zone lattice, the field operator is expanded as follows:
\begin{equation}
\psi({\bf r}) = \sum_i \phi({\bf r} - {\bf r}_i) a^{ }_i \; ,
\label{psi}
\end{equation} 
where $\phi$ is the Wannier function. We thus finally
have
\begin{equation}
n_{\bf k} = | \phi({\bf k})|^2 \sum_{i,j} 
e^{i{\bf k}({\bf r}_i-{\bf r}_j )} \rho_{ij} \; ,
\label{nk}
\end{equation} 
where
\begin{equation}
\rho_{ij} = \langle a^{\dagger}_i 
a^{ }_j \rangle \; ,
\label{rho}
\end{equation}
and $\phi({\bf k})$ is the Fourier transform of
$\phi({\bf r})$.
From Eq.~(\ref{nk}) it is seen that up to a trivial reweighting factor
$|\phi({\bf k})|^2$ the distribution is a periodic function in the 
reciprocal lattice. Thus without loss of generality we may restrict ourselves 
to the first Brillouin zone. Actually, $\phi({\bf k})$ has
nothing to do with the Bose-Hubbard model, being a 
non-universal property of the lattice cite potential; in what follows
we will ignore this function altogether by formally setting it to unity.

Having calculated $\rho_{ij}$ with the Worm algorithm \cite{Worm}, we readily 
obtain $n_{\bf k}$ using Eq.~(\ref{nk}); the results are presented in Figs.~2(a-f). 
In Fig.~2(a) we see a typical picture for the strongly 
correlated superfluid phase, characterized by a single, narrow peak at small
momenta. When a domain of the Mott-insulating phase appears in the 
center of the trap (where the on-site filling is close to unity),
a pronounced fine structure develops in Fig.~2(b). We associate this structure
with the shell-type form of the condensate wave-function. 
To prove the point we model the situation with the pure-condensate density matrix  
$\rho ({\bf r},{\bf r}')=\Psi_0^* ({\bf r}) \Psi_0({\bf r})$, where
the condensate wave-function $\Psi_0({\bf r})$ has the shell-type 
form with the shell radius $l$.
The presence of the Mott insulator is taken into account through 
the suppression of the $\Psi_0({\bf r})$ in the center. The Fourier 
transform of such $\Psi_0({\bf r})$  is alternating in sign, 
with the half-period in $k$ related to the shell radius as 
$k \sim \pi /l$. Thus in the pure condensate we would see exact zeros
in $n_{\bf k}$ with the typical separation between them $\sim \pi /l$. 
Surprisingly, this naive model works extremely well and adequately
describes the case of the realistic strongly correlated system close to
the phase transition (in Fig.~2(b) the coupling $U/t=32$ is close 
to the critical value estimated in Refs.~\cite{Sheshadri,FM}).     
We consider the appearance of the satellite peaks as a clear 
signature of the Mott-insulator transition in the center of the trap. 
\end{multicols}

\begin{multicols}{0}
\onecolumn
\vspace*{-1.0cm}
\begin{figure}
\begin{center}
\epsfxsize=0.33\textwidth
\hspace*{0.0cm} \epsfbox{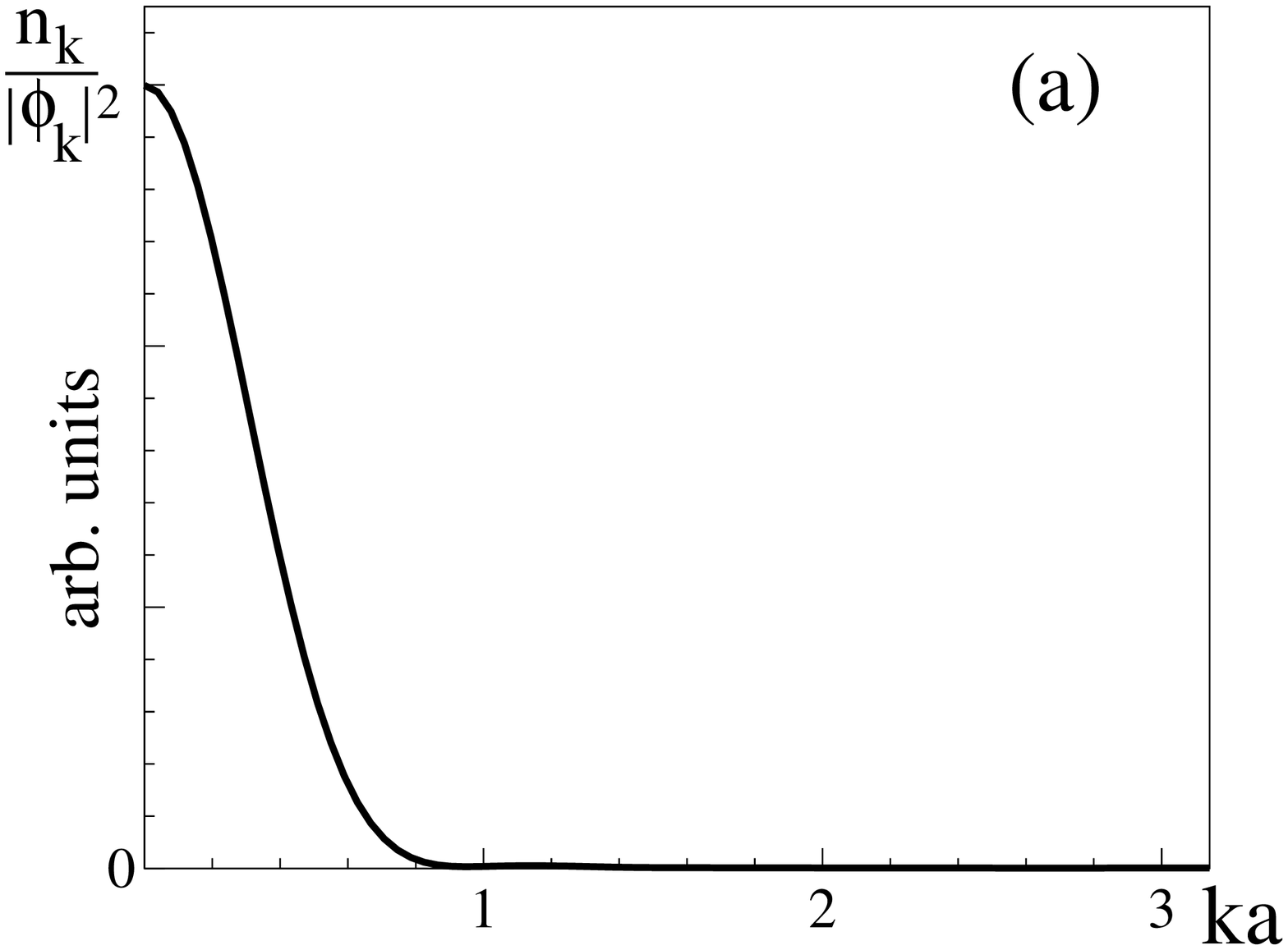}
\epsfxsize=0.33\textwidth
\hspace*{0.0cm} \epsfbox{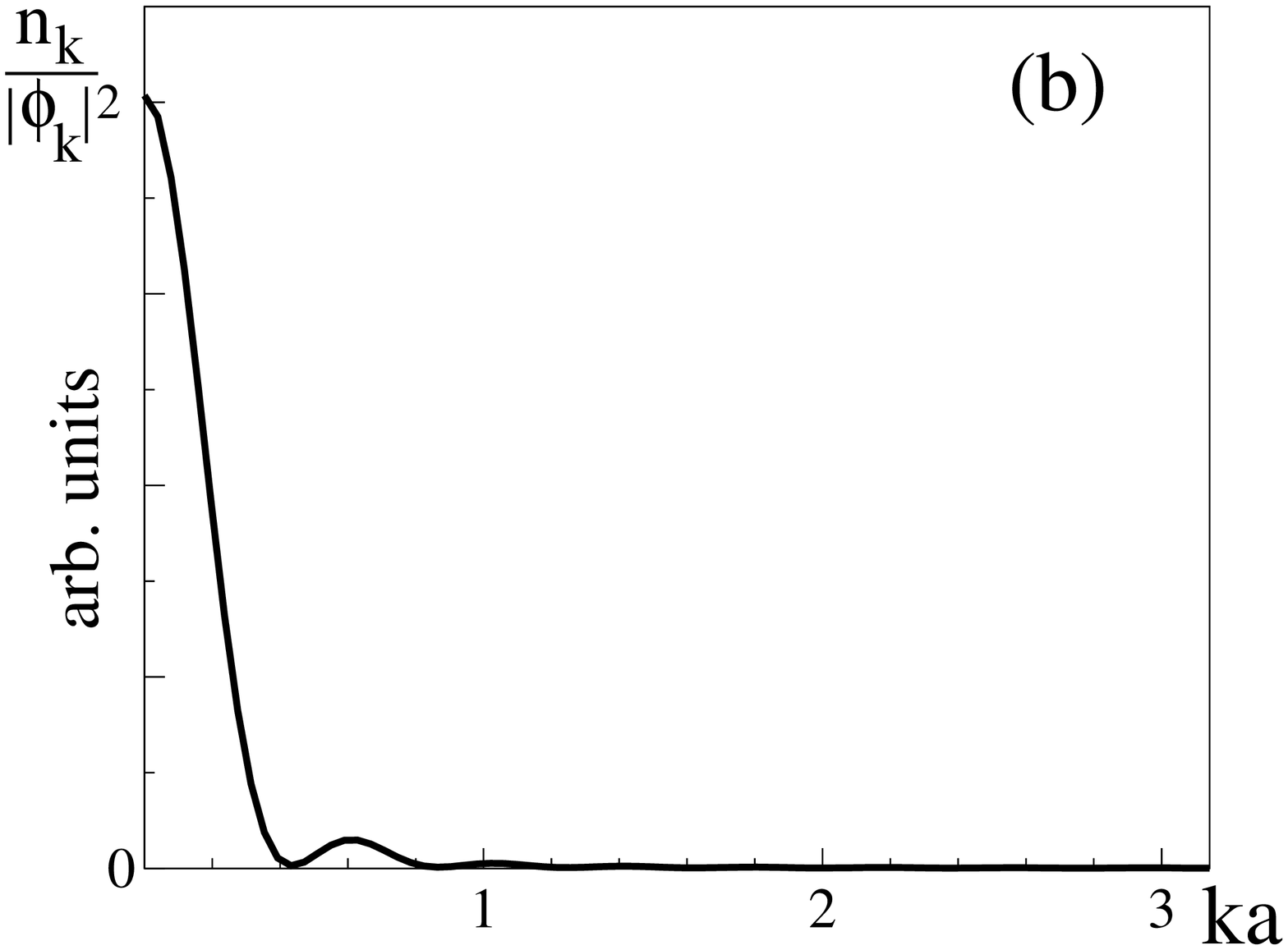}
\epsfxsize=0.33\textwidth
\hspace*{0.0cm} \epsfbox{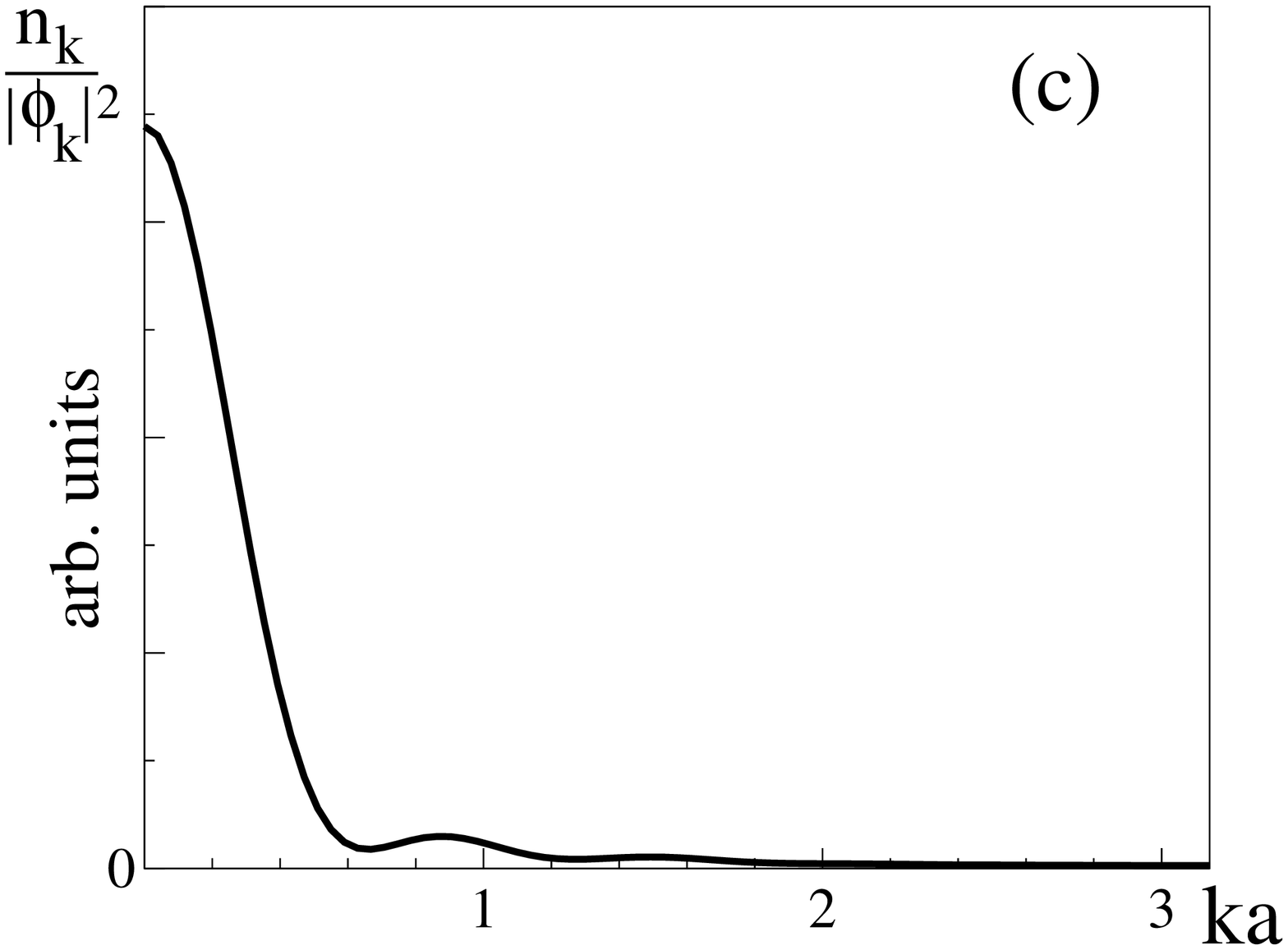} 

\vspace*{-2.5cm}
\epsfxsize=0.33\textwidth
\hspace*{0.0cm} \epsfbox{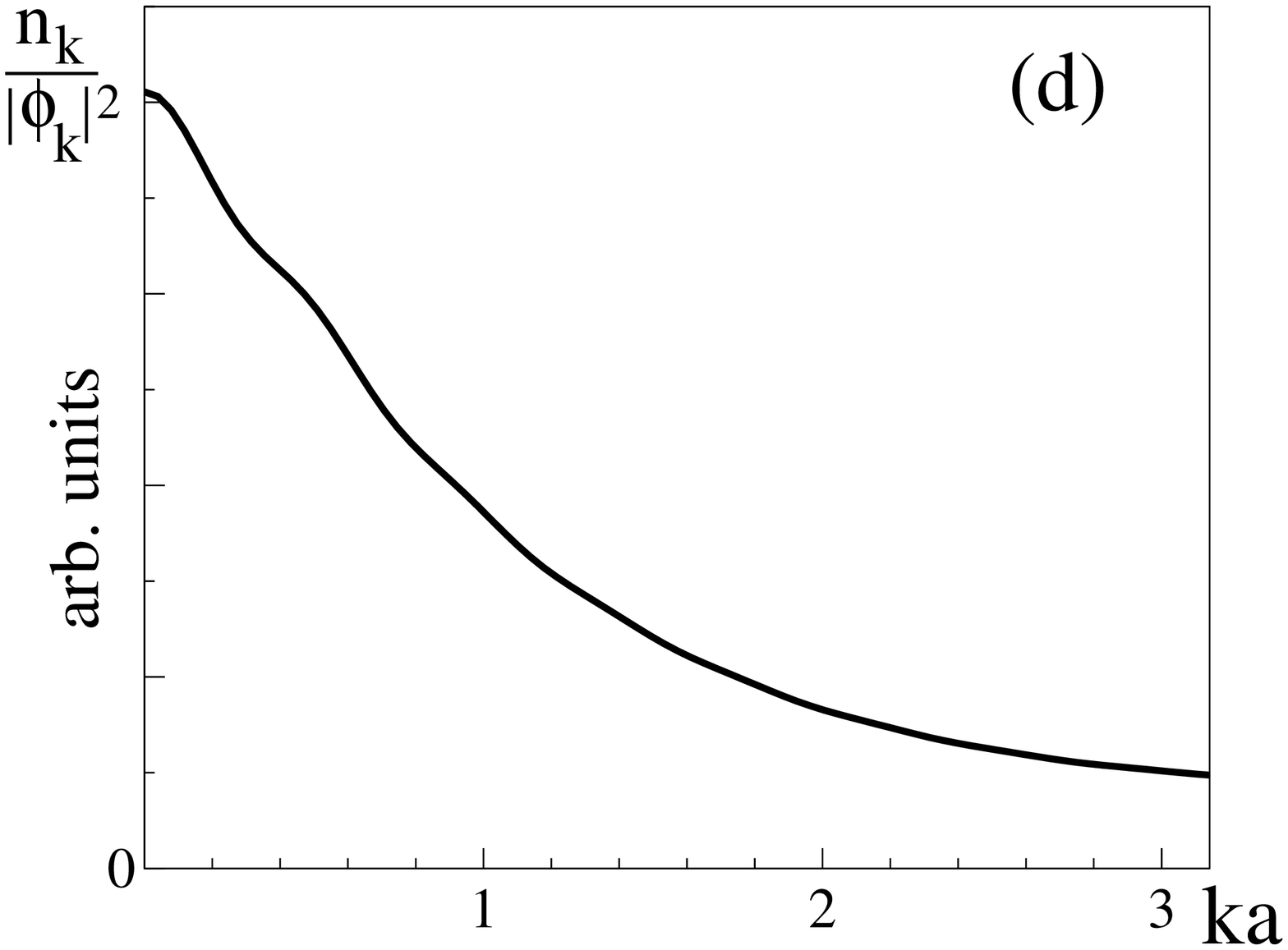}
\epsfxsize=0.33\textwidth
\hspace*{0.0cm} \epsfbox{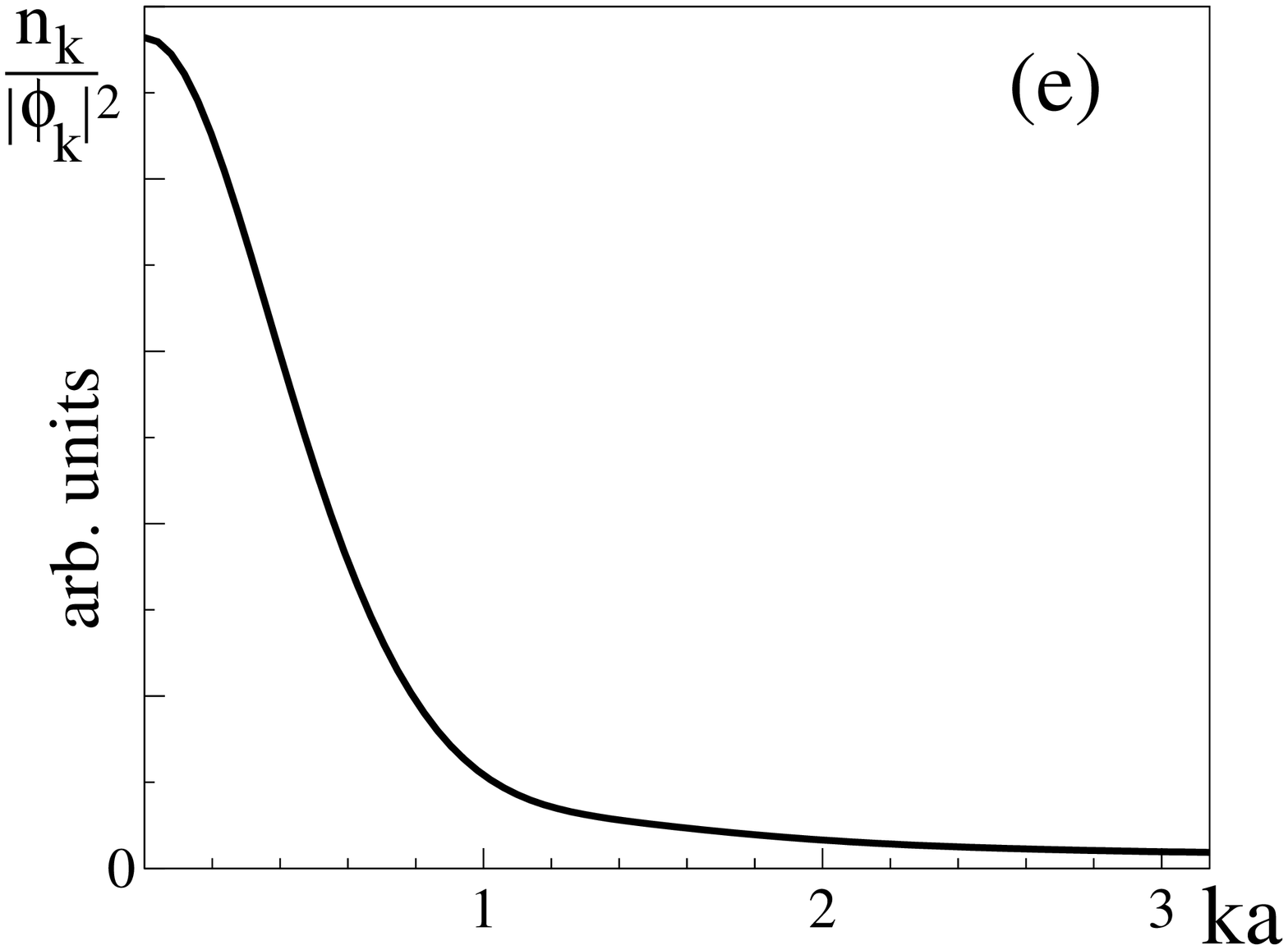}
\epsfxsize=0.33\textwidth
\hspace*{0.0cm} \epsfbox{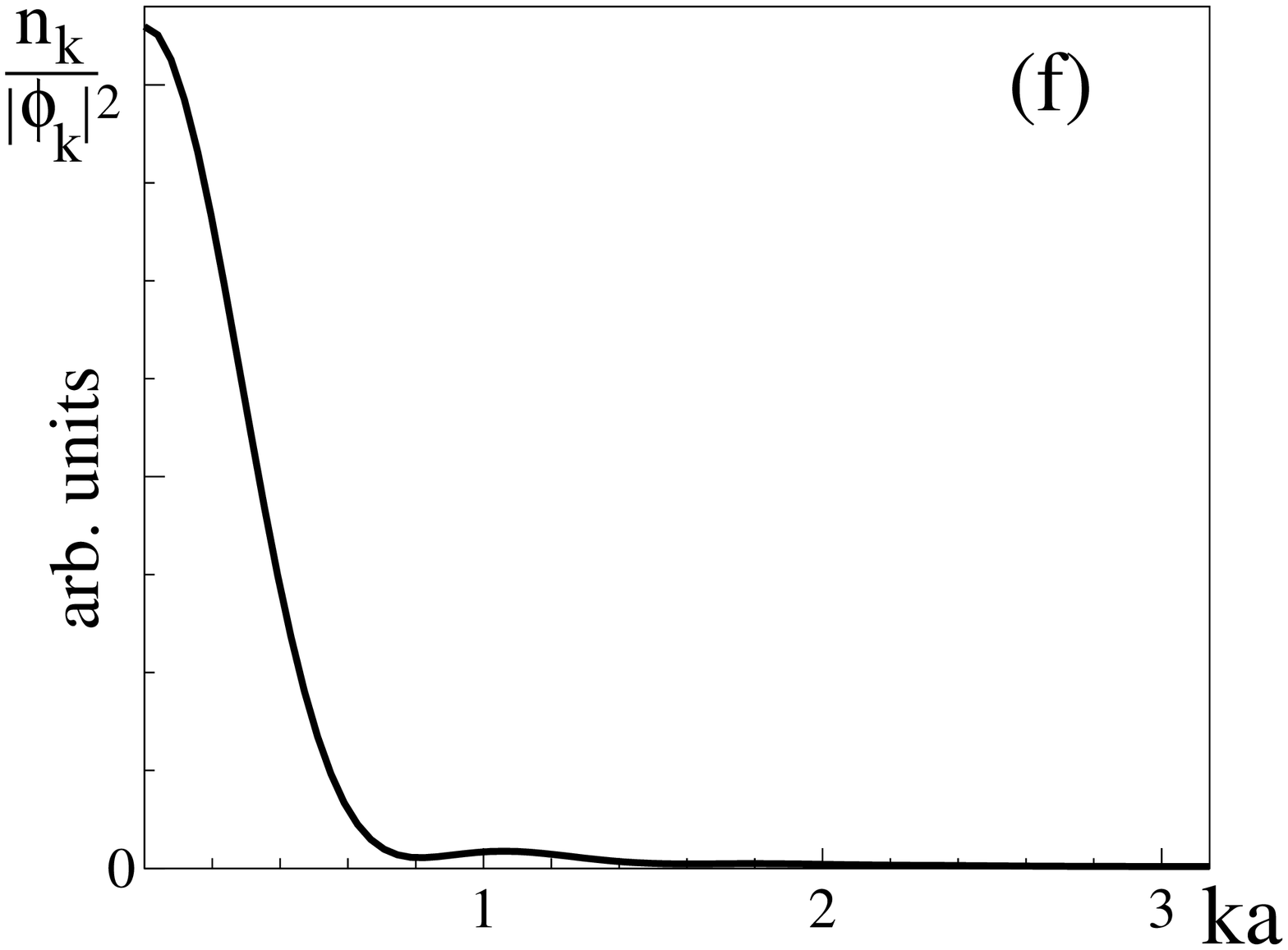}
\end{center}
\vspace*{-1.2cm}
\caption{The $n_k$ distributions in the first Brillouin zone 
in the $(0,0,1)$ direction derived from the single-particle 
density matrices for systems shown in Figs.~1(a-f).}
 
\end{figure}
\end{multicols}

\begin{multicols}{2} 

In Fig.~2(c) the coupling strength is significantly increased to $U/t=80$,
but the shell-type structure of the superfluid phase, and the corresponding
fine-structure in $n_{\bf k}$, is still present. 
The crucial difference between Fig.~2(c) and Fig.~2(d) 
is in the suppression (almost complete) of the superfluid 
fraction. Now the distribution $n_{\bf k}$ has only the central peak and 
is extended towards large momenta, as expected for the system deep in 
the insulating phase. Still, it is not flat (!), which tells us about
large off-diagonal correlations between the nearest lattice cites 
even for $U/t$ as large as  $\sim  2.5 \, (U/t)_c$.   

Fig.~2(e) is similar in physics to the case (a), except for the 
large-momentum tail due to the Mott-insulator shell. Finally, in Fig.~2(f)
we again see the fine structure of satellite peaks reflecting the 
appearance of the Mott-insulator phase in the center of the trap
and the corresponding superfluid phase shell [in close resemblance 
with Fig.~2(b)].  

We note, that momentum distributions $n_{\bf k}$ presented above,
may be observed experimentally if atoms are collimated out of the 
expanding cloud so that the distribution in a given direction is 
photographed. In the current setup, the absorption images of the three
dimensional distribution are taken along two orthogonal axis. This procedure
reveals only the integral 
$N (k_x,k_y) \propto \int_{-\infty}^{\infty} dk_z n({\bf k})$. 
It is easy to see that integration effectively erases fine-structure features
of $n ({\bf k})$---although peaks do not disappear completely, they now
show up as shoulders in $N (k_x,k_y)$.

Finally, we would like to discuss the problem of the repulsive interaction
between particles during the initial period of their free expansion.
Obviously, the interpretation of the final absorption pattern in terms
of the {\it initial} momentum distribution is valid only if the effect
of this interaction is small enough. Meanwhile, given realistic experimental
parameters, this turns out to be the case only for rather moderate
system sizes. The criterion for neglecting the effect of interparticle
repulsion is 
\begin{equation}
E_{\rm kin} / E_{\rm pot} \gg 1 \; ,
\label{cond1}
\end{equation}
where $E_{\rm kin}$ and $E_{\rm pot}$ are, respectively, the kinetic and potential
energy per particle in the most fragile low-momentum part of the distribution
$n_{\bf k}$ and at the most dangerous period of free evolution at the end of 
the restructuring period, when the ``discrete" distribution of density 
transforms into the ``continuous" spatial distribution with the typical size 
of order of the original system size (plus the corresponding replicas 
in higher Brillouin zones).
For the potential energy we have $E_{\rm pot} \sim n U_0$, where $n$ is the
continuous number density, $U_0=4\pi \hbar^2 a_s/m$, $a_s$ is the 
$s$-scattering length,
and $m$ is the atom mass. Recalling that the lattice filling factor is of 
order unity, we can estimate $n \sim 1/a^3$. The lowest kinetic energy is
associated with the spatial distribution of the condensate. Estimating
$E_{\rm kin} \sim  \pi^2 \hbar^2/m a^2 L^2$, where integer $L$ stands for
the typical size of the superfluid component in units of lattice constants,
we arrive at a simple requirement
\begin{equation}
(a/a_s) \, L^{-2} \gg 1 \; .
\label{cond2}
\end{equation}
In the experiment of Ref.~\cite{Greiner}, the ratio $a/a_s$ is of order 
$10^2$. Hence, we are restricted to $L<10$, that is to 
typical system sizes of our present simulation. 
Note also, that the condition (\ref{cond1}) is much
easier to satisfy for the atomic cloud in the {\it second} 
Brillouin-zone peak,  where the spatial density is significantly 
suppressed \cite{Greiner}.

Summarizing, we presented a simulation of the ground state properties
of ultracold atoms in an optical
lattice with confining external potential, in the regime where the 
Mott-insulator and superfluid phases coexist. We have demonstrated that when
the insulator domain in the center of the trap is surrounded by the superfluid
component, the global momentum distribution of particles features satellite peaks.
This picture can be employed by the experiment as an unambiguous evidence of the
Mott transition. We do not see other features of the momentum distribution
that could be associated with the Mott transition: unless the ratio $U/t$ is
not much larger than the critical one, the momentum distribution in the 
reciprocal lattice still has a peaked form reflecting strong 
local off-diagonal correlations.

This work was supported by the National Science Foundation 
under Grant DMR-0071767. BVS acknowledges a support 
from the Russian Foundation for
Basic Research (RFBR) under Grant 01-02-16508 and from the Netherlands
Organization for Scientific Research (NWO). 
VAK acknowledges a support from RFBR under Grant 00-02-17803.
We acknowledge
helpful discussions with H. Monien and Yu. Kagan.

\end{multicols}

\end{document}